\documentclass[prb,twocolumn,reprint,amsmath,amssymb,showpacs,floatfix,superscriptaddress]{revtex4}
\usepackage{color}
\usepackage{graphicx}

\usepackage[sort&compress]{natbib}
\usepackage{color}
\usepackage{soul}

\usepackage{hyperref}
\hypersetup{
bookmarks=false,
pdftitle={Microwave photonics}
anchorcolor=blue,
colorlinks=true,
citecolor=blue,
urlcolor=blue,
linkcolor=blue}
\usepackage{xcolor}

\usepackage[normalem]{ulem}

\begin{document}

\title{Tunable coupling engineering between superconducting resonators:\\
from sidebands to effective gauge fields}

\author{Borja Peropadre}
\address{Instituto de F{\'i}sica Fundamental, IFF-CSIC, Calle Serrano 113b, E-28006 Madrid}
\author{David Zueco}
\address{Instituto de Ciencia de Materiales de Aragï¿½\'on y
  Departamento de F\'{\i}sica de la Materia Condensada CSIC - Universidad de
  Zaragoza C/ Pedro Cerbuna 12, 50009 Zaragoza (Spain)}
\address{Fundaci\'on ARAID, Paseo Mar\'ia Agust\'in 36, 50004 Zaragoza, Spain}
\author{Friedrich Wulschner}
\address{Walther-Mei{\ss}ner-Institut, Bayerische Akademie der Wissenschaften, 85748 Garching (Germany)}
\author{Frank Deppe}
\address{Walther-Mei{\ss}ner-Institut, Bayerische Akademie der Wissenschaften, 85748 Garching (Germany)}
\address{Technische Universit\"at M\"unchen, Physik Department,
James-Franck-Str., D-85748 Garching, Germany}
\author{Achim Marx}
\address{Walther-Mei{\ss}ner-Institut, Bayerische Akademie der Wissenschaften, 85748 Garching (Germany)}
\author{Rudolf Gross}
\address{Walther-Mei{\ss}ner-Institut, Bayerische Akademie der Wissenschaften, 85748 Garching (Germany)}
\address{Technische Universit\"at M\"unchen, Physik Department,
James-Franck-Str., D-85748 Garching, Germany}
\date{\today}
\author{Juan Jos\'e Garc{\'\i}a-Ripoll}
\address{Instituto de F{\'i}sica Fundamental, IFF-CSIC, Calle Serrano 113b, E-28006 Madrid}

\begin{abstract}
In this work we show that a tunable coupling between microwave resonators can be engineered by means of simple Josephson junctions circuits, such as dc- and rf-SQUIDs.  We show that by controlling the time dependence of the coupling it is possible to switch on and off and modulate the cross-talk, boost the interaction towards the ultrastrong regime, as well as to engineer red and blue sideband couplings, nonlinear photon hopping and classical gauge fields. We discuss how these dynamically tunable superconducting circuits enable key applications in the fields of all optical quantum computing, continuous variable quantum information and quantum simulation --- all within the reach of state of the art in circuit-QED experiments. 
\end{abstract}

\maketitle

\section{Introduction}
\label{sec:introduction}


The field of circuit quantum electrodynamics (circuit-QED) studies the interaction between artificial atoms and artificial photons~\cite{blais04,wallraff04}, both of them implemented with the same technology: superconducting circuits cooled to millikelvin temperatures. A key feature of these systems is that, based on the same microscopic model, both the photonic degrees of freedom and the artificial atoms have similar energy scales and may interact very strongly. Hence, they show effects which are beyond those explored in the optical domain. A paradigmatic example is the failure of the rotating wave approximation when the qubit-photon coupling approaches the qubit and photon energies~\cite{niemczyk10,forn-diaz10}. Aside from the development of qubits and the control of their interactions~\cite{Hime2006, Niskanen2007, VanderPloeg2007}, circuit-QED has recently started to focus on photons themselves, mostly in the context of two different experimental configurations. In the first type of setups, cavities are replaced with open transmission lines and propagating microwave photons that move and interact with localized qubits. This allows us to study one-dimensional artificial QED, atom-light interaction~\cite{astafiev10}, electromagnetically induced transparency~\cite{abdumalikov10}, causality~\cite{sabin11}, quantum metamaterials \cite{Rakhmanov2008, Hutter2011, Zueco2012} and to implement photodetectors~\cite{romero09,peropadre11} and routers~\cite{hoi11}. The other type of setups is based on polariton physics~\cite{hartmann06,angelakis07}: by coupling multiple cavity-qubit systems it is possible to build lattices on which dressed photons hop and interact, either attractively or repulsively, implementing Hubbard type models or spin Hamiltonians~\cite{leib10, Leib12}. This gives rise to well known models, such as a Tonks-Girardeau gas~\cite{carusotto09}, however the architecture based on superconducting cavities and Josephson junctions also allows for the exploration of new phenomena, such as gauge fields and frustration~\cite{nunnenkamp11}.

In this work we revisit the architecture of coupled superconducting cavities, designing a tunable coupling between nearest-neighbor resonators. This represents a major breakthrough for this type of systems, because the dynamical tunability of the resonator coupling makes it possible to engineer a huge variety of \textit{photon-photon} interactions: from red and blue sidebands to gauge fields, passing through correlated photon hopping and Kerr nonlinearities, or simply canceling the usual cross-talk between resonators. All these approaches are based on simple superconducting circuit elements, such as dc- and rf-SQUIDs.

The setup that we have in mind consists of an array of linear resonators, connected through different types of Josephson junction (JJ) circuits [cf. figure~\ref{fig:array}]. The fixed circuit structure is associated with static, geometry dependent capacitive and inductive couplings between the resonators, while those related to the Josephson junction circuits can be tuned by an applied magnetic flux. Most notably, the total coupling strength can be reduced and even completely suppressed as the coupling due to the Josephson junction circuits can have opposite sign. 

 Unlike previous proposals that rely on dispersive coupling via qubits~\cite{Mariantoni2008, Reuther2010} or through circulator-like elements~\cite{chirolli10}, the coupling scheme proposed here has the potential to be stronger --- the coupling elements are galvanically coupled, as in Ref.~[\onlinecite{peropadre10}] ---, it is more robust against external perturbations (charge noise) and the geometry of the lattice is not tied to the coupling element. The coupling elements can be operated in two ways: i) with a stationary configuration of magnetic fields that determines the associated coupling matrix between oscillator modes or ii) with a periodic multicolor driving that allows for engineering arbitrary sideband interactions, $\eta_1 a^\dagger b + \eta_2 a b + \mathrm{H.c.}$ between any two resonators, $a$ and $b$, with adjustable coupling strengths $\eta_{1,2}$. Moreover, our design achieves tunability in a regime in which the couplings are strong or ultra-strong --- larger than the corresponding decay rates or comparable to the cavity frequencies, respectively ---, regardless of other elements that may coexist with the cavities, such as qubits or magnetic impurities.

As potential applications of this work we would like to address two fields. The first one has been sketched above: by tuning the coupling between different cavities it is possible to tune the lattice topology, the coupling strength and even the phase of the hopping terms in polariton arrays. This nicely complements existing proposals which show how to tune the photon nonlinearity by manipulating the qubit inside the cavity~\cite{leib10} and gives access to effective gauge fields without relying on fragile coupling elements~\cite{nunnenkamp11}. The second type of application points along the line of quantum information and the manipulation of continuous variable states. By means of the coupling circuits in this toolbox one may implement any nearest neighbor quadratic Hamiltonian with any time dependence and geometry, as far as it is embeddable in a 2D manyfold. This can be used to implement interesting states, such as two-dimensional continuous variable Gaussian states~\cite{menicucci06}, whose tomography could be supplemented by embedded qubits~\cite{hofheinz09} or moving probes~\cite{shanks12}.

The paper is organized as follows.  In the first part [Sect.~\ref{sec:inductive}] we will study two superconducting resonators that are close together and subject to a mutual inductive and capacitive interaction. Using the Lagrangian quantization, we will show that, both in the weak and strong coupling regime, the geometric crosstalk gives rise to a constant beam-splitter type interaction. In the second part of this work (Sect.~\ref{sec:effective}) we propose two quantum circuits that dynamically tune the inductive coupling between the resonators. The first one is a SQUID which is galvanically coupled to two resonators. The second one uses instead two coupling wires, creating an interference device between resonators. We will discuss analytically both models, demonstrating that they can tune and switch off the overall resonator-resonator coupling. In Sec.~\ref{subsec:strength} we study the validity of our designs under realistic experimental conditions, estimating the coupling strengths that can be attained in current experiments. In Sect.~\ref{sec:sidebands} we consider the situation of a time-dependent resonator coupling. We show that a periodic modulation of the coupling makes it possible to engineer sidebands in a non-perturbative fashion, controlling the strength and phase of both the rotating and counterrotating terms. Finally, in Sec.~\ref{sec:conclusions} we summarize our results and suggest a large set of potential applications, ranging from quantum information to quantum simulation.

\section{Static coupling}
\label{sec:inductive}
In this section we derive the Hamiltonian that rules the dynamics of two coupled superconducting strip line resonators, and give a general expression for the different coupling constants that arise from the model. Firstly we will consider the simplest case of coupling, caused by the mutual inductance and mutual capacitance due to the spatial proximity of the resonators. Since the coupling is time independent and determined by the detailed spatial arrangement of the resonators, we refer to it as static geometric coupling. The discussion of this interaction is done for a particular configuration of parallel resonators, but the objective is just to exemplify how this coupling manifests itself as a beam-splitter interaction.


\begin{figure}
\includegraphics[width=\columnwidth]{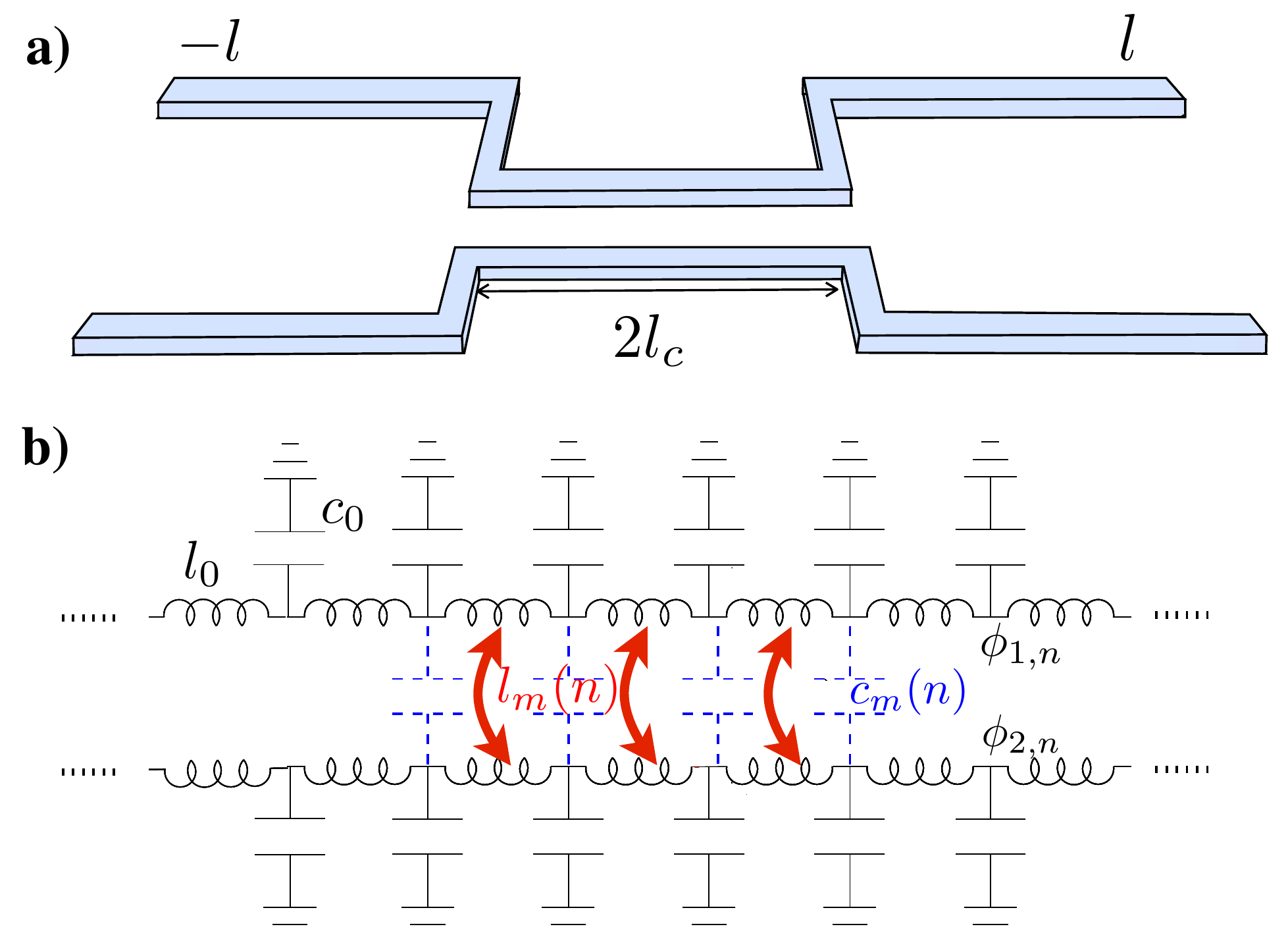}
\caption{a) Sketch of the geometrical arrangement of the two coupled superconducting stripline resonators of length $2l$. A finite interaction is present only in the coupling region of length $2l_c$ and is negligible elsewhere. b) Schematics for the lumped circuit equivalent. We explicitly draw the mutual capacitances (dashed blue lines) and the mutual inductive coupling (red arrows). The node flux $\phi_n$ is also indicated.}
\label{fig:lumped}
\end{figure}
 
Consider two superconducting stripline resonators of length $2l$, as it is depicted in Fig.~\ref{fig:lumped}a.  In this particular layout, the coupling occurs mainly within a middle section of length $2l_c$, where the resonators approach each other. Assuming that the crosstalk is given by the mutual inductance $l_m$ and mutual capacitance $c_m$ induced in this middle region~\cite{Clayton1994, Clayton2002}, we can write down the following Lagrangian density:
\begin{equation}
\label{density_lag}
\mathcal{L}=
\sum_{i,j}\int_{-l}^{l} \left[ \frac{{\hat c}_{ij}}{2} \dot{\phi}_i \dot{\phi}_j
-\left(\frac{1}{2 \hat{l}}\right)_{ij} \partial_x{\phi_i}\partial_x{\phi_j} \right] dx,
\end{equation}
where both the flux fields, $\phi_i(x)$, the capacitance, $\hat{c}$, and inductance matrices, $\hat{l}$, depend on the position along the transmission line
\begin{align}
\label{cl_matrices}
\hat c
&=
c_0(x) + c_m(x)
( \openone -\sigma_x)
 ,
\\ 
\hat l
&=
l_0(x) + l_m (x) \sigma_x
\; ,
\end{align}
where $\sigma_x$ is the Pauli matrix.  A full derivation of~(\ref{density_lag}) can be obtained from the lumped circuit equivalent of the strip lines [Fig.~\ref{fig:lumped}b] and it is thoroughly discussed in App.~\ref{app:A}.

For the sake of simplicity, we will consider the capacitance and inductance per unit length of each line, $c_0(x)$ and $l_0(x)$, to be constant, and use piecewise constant functions for the mutual inductance and capacitance
\begin{equation}
\label{c_m}
c_m(x)= 
\left \{
\begin{array}{cc}
c_m & |x| <l_c
\\
0& {\rm otherwise}
\end{array}
\right .,\;
l_m(x)= 
\left \{
\begin{array}{cc}
l_m & |x| < l_c
\\
0& {\rm otherwise}
\end{array}
\right . .
\end{equation}

We derive a normal mode expansion for the flux $\phi_j(x,t) = \sum_n q_{j,n}(t) u_n(x)$ in each resonator $j=1,2$. In what follows we restrict ourselves to the fundamental mode of each resonator
with frequency $\omega_0$ and total capacitance $C_r = \int_{-l}^{l} c_0(x) dx$. Within this subspace and mode expansion, the interaction term gives rise to off-diagonal terms, as expected from an interaction between two cavities, but also diagonal terms that induce a renormalization (dressing) of the oscillator frequencies. This dressed resonance frequency is
\begin{eqnarray}
\omega = \omega_0 \sqrt{
\frac{1+C}{1 + 2 C} \left ( 
1 + \frac{1}{\nu}\frac{L^2}{1 - L^2}
\right )
},
\end{eqnarray}
expressed in terms of two \textit{overlap} integrals
\begin{equation}
\Delta_1 = \int_{-l_c}^{l_c}  u_0(x)^2 \mathrm{d}x,\;
\Delta_2 = \int_{-l_c}^{l_c} \left[\partial_x u_0(x)\right]^2 \mathrm{d}x,
\end{equation}
where $C = c_m \Delta_1/C_r$, $L = l_m/l_0$, and  $\nu = {\omega_0^2 C_r l_{0}}/{\Delta_2}$ is a geometric factor.


We finally proceed with the quantization of this model, introducing the oscillator length
$a_0 = \sqrt{\hbar(1+C)/C_r \omega(1+2C)}$.   We express the phase space operators in terms of the Fock operators,
$q_j = a_0 (a_j + a_j^\dagger)/\sqrt{2}$ and $p_j = i \hbar (a^\dagger_j - a_j)/\sqrt{2}a_0$.
This leads to,
\begin{eqnarray}
\label{quant_ham}
H &=& 
\hbar\omega  \sum_{j=1,2} a_j^\dag{a}_j^{} - \hbar g_{c}
({a}_1^\dag-{a}_1^{}) ({a}_2^\dag-{a}_2^{}) \\
&-& \hbar g_{i} ({a}_1^\dag+{a}_1^{})({a}_2^\dag+{a}_2^{}).\nonumber
\end{eqnarray}
The coupling constants $g_c$ and $g_i$ account for the static capacitive and inductive contributions to the coupling, respectively,
\begin{align}
\label{gc}
   g_{c} &= \frac{\omega_0}{2} \sqrt{\frac{C^2}{(1+C)(1 + 2 C)} \left ( 
1 + \frac{1}{\nu}\frac{C^2}{1 - C^2}
\right )},
\\
\label{gi}
   g_{i} &= \frac{\omega_0}{2}
\frac{1}{\nu}\frac{L}{1-L^2} \sqrt{\frac{1+C}{1 + 2 C} \frac{1}{ 
1 + \frac{1}{\nu}\frac{L^2}{1 - L^2}}}.
\end{align}

The usual limits in quantum optics correspond to the weak coupling and strong coupling regimes. In both of them $g_{{c,i}}/\omega_0\ll 1$, so that the frequency renormalization becomes negligible (provided that $C,L\ll1$). We can then invoke the Rotating Wave Approximation (RWA) and transform (\ref{quant_ham}) to the beam splitter model,

\begin{equation}
 H \simeq \hbar \omega_0\sum_{j=1,2}  a^\dagger_j a_j - \hbar(g_i + g_c)({a}_1^\dag{a}_2^{}\,{+}\,{a}_2^\dag{a}_1^{}).
\label{eq:jc}
\end{equation}
Note how the resulting Hamiltonian can be interpreted as an exchange or \textit{hopping} of excitations between modes, similar to optical lattice and tight-binding models. 

This type of static geometric coupling is implicit in the experimental configurations of coupled cavity models~\cite{hartmann06,angelakis07}, though previous designs have inclined to consider a capacitive coupling taking place at the electric field nodes (current anti-nodes) of the resonator~\cite{leib10, koch10, nunnenkamp11}, sometimes enhanced by an additional JJ circuit~\cite{koch10,nunnenkamp11}.


\section{Tunable coupling}
\label{sec:effective}

\begin{figure}[t!]
  \centering
 \includegraphics[width=\linewidth]{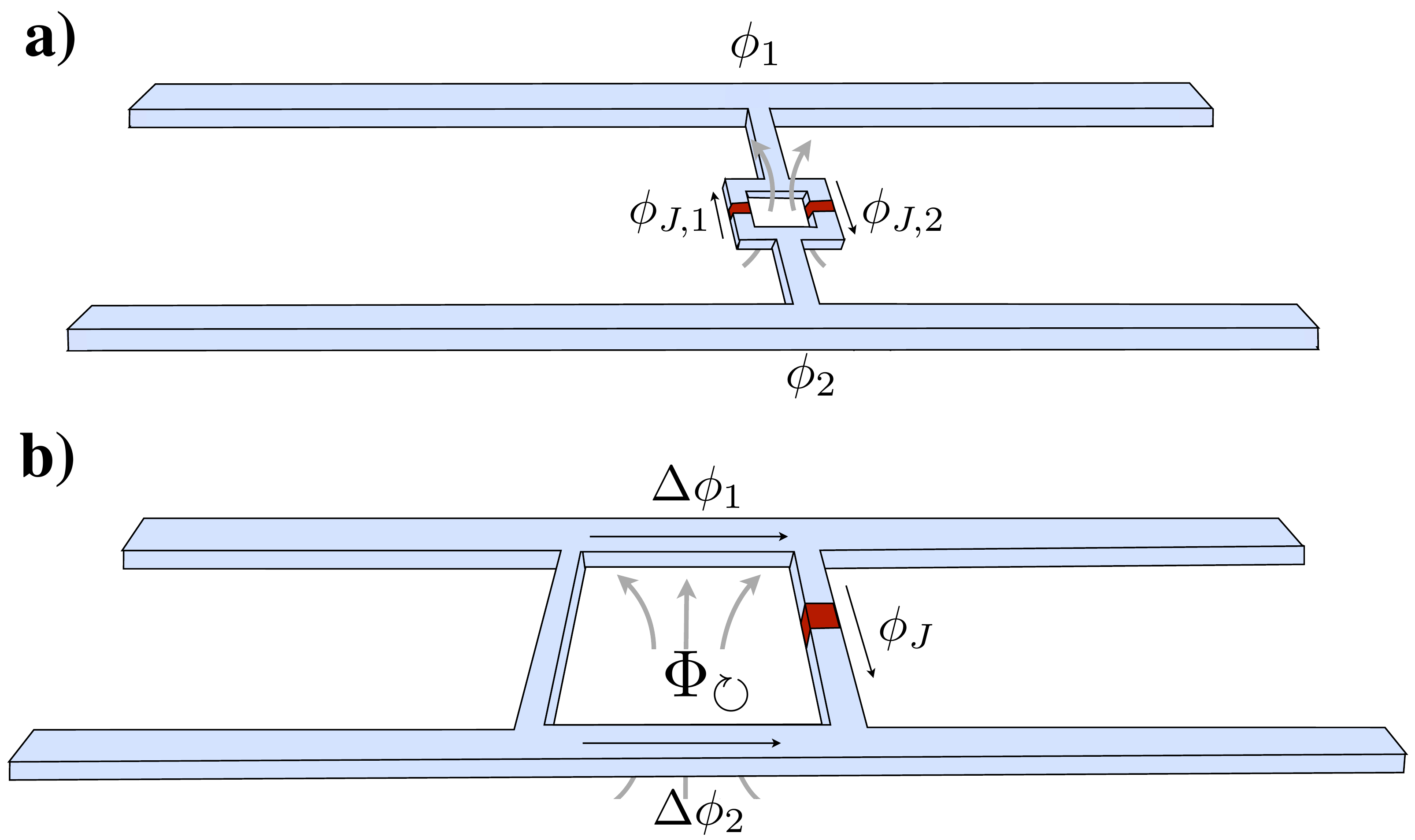}
  \caption{\small{(color online) Architectures leading to a tunable microwave beam splitter. a) A dc SQUID (superconducting loop interescted by two Josephson junctions) mediates the coupling between two stripline resonators. The established pointlike contact between the resonators takes the coupling to the ultrastrong domain. b) A superconducting ring intersected by a Josephson junction now partially shares its branches with the cavities, improving the switching capability.}}
  \label{setup_beam}
\end{figure}

In this section we study alternative mechanisms for coupling two or more linear resonators. On the one hand we aim at a larger coupling strength, and on the other hand we wish to achieve real-time tunability of the couplings. For both goals it will be advantageous to rely on inductive rather than capacitive coupling. First of all, the inductive coupling realized by JJs and loops intersected by JJs (e.g. SQUIDs) can be tuned by an applied magnetic field varying the magnetic flux threading the JJs or loops.  Second, and equally important, inductive interactions can be enhanced, profiting both from the kinetic inductance of thin superconducting films and from embedded junctions working in the linear regime~\cite{bourassa09}. Based on the two previous ideas, we envision the two coupling elements sketched in Fig.~\ref{setup_beam}. We will study analytically these designs, deriving  expressions for the effective interactions and coupling strengths.

\subsection {SQUID as a coupler}
\label{sect:squid-coupler}

Given the large inductance provided by the Josephson junctions, one could naively think of connecting both cavities with a superconducting wire interrupted by a Josephson junction. In doing so one would achieve a static ultrastrong coupling. However, for tuning the Josephson inductance of a single junction we have to generate a magnetic flux of the order of a flux quantum threading the junction area. Due to the small junction area unpractically large magnetic fields would be required. Fortunately, we can design a much better tunability by using a superconducting quantum interference device (SQUID) configuration, as depicted in Figure~\ref{setup_beam}a. Since the SQUID loop area is much larger than the junction area, much smaller control fields are required. Note that even though the use of dc-SQUIDs as tunable couplers was already suggested for flux qubits~\cite{grajcar06}, our setup works with continuous variables and has subtle differences that need to be discussed below.

A short line with the SQUID represents a small contribution to the original Lagrangian density~(\ref{density_lag}), that is ${\mathcal L_t}={\mathcal L}+{\mathcal L_\textrm{SQUID}}$ with%
\begin{equation}
\label{LSQUID}
\mathcal{L}_{\textrm{SQUID}}=\sum_{k=1}^2\frac{C_{J,k}}{2}\dot{\phi}_{J,k}^2+E_{J,k}\cos\left({\frac{2\pi\phi_{J,k}}{\Phi_0}}\right),
\end{equation}
where, $\phi_{J,k}$ represent the flux differences along the junctions $k=1,2$. We use fluxoid quantization along the SQUID loop, $\phi_{J,1}+\phi_{J,2}+\Phi_\circlearrowright=n\Phi_0$, to express the Lagrangian in terms of the variables $\phi_{\pm}=\frac{1}{2}(\phi_{J,1}\pm\phi_{J,2})$ and the total flux enclosed by the loop, $\Phi_\circlearrowright$. For simplicity, we assume $\Phi_\circlearrowright \simeq \Phi_{\rm ext}$, that is, we are neglecting the additional flux generated by the circulating loop current. This is equivalent to restricting our discussion to screening parameters $\beta_L =2\pi L I_c/\Phi_0 \ll 1$ as discussed in more detail below. Here, $L$ is the loop inductance and $I_c$ the critical current of the Josephson junctions. If the SQUID is symmetric, $C_{J,1}=C_{J,2}$ and $E_{J,1}=E_{J,2}$, the coupling becomes $E_{\textrm{eff}}\cos\left({{\pi\phi_{-}}/{\Phi_0}}\right)$, with an effective Josephson coupling energy that depends on the flux threading the SQUID loop,
\begin{equation}
\label{Eeff}
E_{\textrm{eff}}=2E_J \cos\left({\pi\Phi_{\circlearrowright}/\Phi_0}\right).
\end{equation}
The voltage-phase relation $\dot{\phi}_{-}=\dot{\phi}_1(x)-\dot{\phi}_2(x)$  allows us to express $\phi_-$ in terms of the voltages at the edges of the connecting wire. In the linear limit of small fluxes, i.e. small photon number [cf. App. \ref{ap:B}], we can write a quadratic coupling between fields
\begin{equation}
  \label{eq:linearized-SQUID}
\mathcal{L}_{\textrm{SQUID}}\simeq
\frac{C_J}{2}\left(\dot{\phi}_1-\dot{\phi}_2\right)^2-\frac{2\pi^2E_{\textrm{eff}}}{\Phi_0^2}(\phi_1-\phi_2)^2,
\end{equation}
which by means of the normal mode decomposition adopts the form of Eq.~(\ref{quant_ham}). The tunability of this
model relies on the fact that the Josephson energy
$E_{\mathrm{eff}}$ is flux-dependent. This implies that
\begin{equation}
  \label{g-dependence}
  g_i = g_i^{\mathrm{static}} + \frac{4\pi^2}{\Phi_0^2}E_J \cos\left({\pi\Phi_{\circlearrowright}/\Phi_0}\right)
\end{equation}
in Eq.~(\ref{eq:jc}) can be changed in magnitude and sign. For an appropiate value of the external flux (close to $\Phi_{\circlearrowright}=\Phi_0/2$ if $|g_c|\ll|g_i|$) we can fully deactivate the coupling.

We finally notice that our setup is robust against small differences in the two junction energies, $E_{J1,2}=E_J(1\pm\varepsilon)$. In this case one can still expand $\phi_{1,2}= \frac{1}{2}\Phi_{\circlearrowright}\pm \phi_-$, linearizing around $\phi_-\simeq 0$ to obtain
\begin{eqnarray}
  \mathcal{L}_{\mathrm{SQUID}} \pm 2 \varepsilon E_J \sin(\pi\Phi_{\circlearrowright}/\Phi_0)
  \frac{2\pi\phi_-}{\Phi_0}.
\end{eqnarray}
The linear term in this equation amounts just to a displacement of the oscillators and does not add up to the total coupling, preserving the tunability of the setup. We will use this idea in the following setup.

\subsection{Superconducting ring coupler}
\label{sect:ring}
The second design is shown in Fig.~\ref{setup_beam}b. It consists of a superconducting ring interrupted by a single Josephson junction. Since both resonators share a branch of the loop that couples them, the Lagrangian acquires new contributions with the kinetic inductance of the superconductor~\cite{bourassa09}. This kinetic coupling can be very strong, while still retaining the switching capability due to the fluxoid quantization inside the loop, similar to previous designs for superconducting qubits~\cite{peropadre10}.

Our derivation is based on two non-essential constraints. The first one is that the loop is small enough to neglect its self inductance ($\beta_L\ll 1$). The second one is that the wire without junction touches the resonators at points where the flux of the coupled modes is zero~\footnote{Typically this happens at the middle point for the fundamental mode. This is equivalent to choosing the phase $\varphi_1 (x_1) = 2\pi \phi_1(x_1)/\Phi_0 = \varphi_2 (x_2) = 2\pi \phi_2(x_2)/\Phi_0 =0$ and neglecting the phase drop along the connecting wires compared to that across the Josephson junction. In this case the fluxoid quantization in the loop reads $\Delta \phi_1 - \Delta \phi_2 + \phi_J + \Phi_\circlearrowright = n \Phi_0$. The former assumption can always be made since the current flowing in the resonator only depends on the spatial derivative of the phase and not on its absolute value. The latter is a good approximation as long as the Josephson inductance is large compared to the kinetic inductance of the wires. This is the case as long as the superconducting wire is not made extremely narrow. } $u_0(x_1)=0$. Under these circumstances the coupling term reads
\begin{equation}
\label{cal_L_t}
{\mathcal L_{JJ}}=\frac{C_J}{2}\dot{\phi}_J^2+E_J\cos\left({\frac{2\pi\phi_J}{\Phi_0}}\right),
\end{equation}
The fluxoid quantization inside the loop, $\Delta\phi_2-\Delta\phi_1+\phi_J=-\Phi_{\circlearrowright}$, allows us to get rid of the flux variable $\phi_{J}$ and rewrite the coupling in terms of the branch fluxes $\Delta \phi_{1,2}$,
\begin{eqnarray}
  \label{cal_L_JJ}
  {\mathcal L_{JJ}}&=&\frac{C_J}{2}
  \left (\Delta{\dot{\phi}_1}-\Delta{\dot{\phi}_2}\right
)^2\\
&+&E_J\cos\left[{\frac{2\pi\left (\Delta{\phi_1}-\Delta \phi_2+\Phi_{\circlearrowright}\right)}{\Phi_0}}\right].\nonumber
\end{eqnarray}

At this point we will repeat the linearization of the cosine, much like in~(\ref{eq:linearized-SQUID}). However, now the Taylor expansion will depend on the external flux, $\Phi_{\circlearrowright}$, producing linear and quadratic contributions of different magnitude. We start with the normal mode decomposition of the branch fluxes and restrict ourselves to the lowest energy modes
\begin{equation}
\label{normalmodes}
\Delta\phi_j=\phi_j(x_2)-\phi_j(x_1)\simeq q^{(j)}_1\partial_xu_n(x)|_{x_1}\Delta x.
\end{equation}
Introducing these terms in the interaction produces the quadratic Lagrangian for the fundamental modes,
\begin{eqnarray}
\label{L_JJ}
L_{JJ}&=&\frac{1}{2}\sum_{j=1,2}\left(\alpha_J\dot{q}_j^2-\beta_Jq_j^2\right)\\
&+&\gamma_J(q_1-q_2)-\alpha_J\dot{q}_1\dot{q}_2+\beta_Jq_1q_2.\nonumber
\end{eqnarray}
The expressions for all coefficients can be computed from first principles
\begin{eqnarray}
&&\alpha_J=C_J\partial_xu_0(x=0)^2\Delta
x^2,\\
&&\beta_J=E_J
 \frac{4\pi^2}{\Phi_0^2}\partial_xu_0(x=0)^2\Delta
 x^2\cos\left({\frac{2\pi\Phi_{\circlearrowright}}{\Phi_0}}\right),\\
&&\gamma_J=E_J\frac{2\pi}{\Phi_0}\partial_xu_0(x=0)\Delta
x\sin\left(\frac{2\pi\Phi_{\circlearrowright}}{\Phi_0}\right).
\end{eqnarray}
Out of these terms, $\gamma_J$ is a linear displacement of the cavity eigenmodes and does not transfer energy. The capacitive and inductive terms, $\alpha_J$ and $\beta_J$, are the only ones that contribute to the inter-cavity coupling, $g_c$ and $g_i$, and to the frequency renormalization. More precisely, we obtain the model~(\ref{quant_ham}) with mode frequency
\begin{equation}
\omega=\omega_0\sqrt{1+\frac{\beta_J}{C_r\omega_0^2}}.
\end{equation}
and coupling strengths
\begin{equation}
\label{coupling}
g_i= g_i^{\mathrm{static}}+\frac{\beta_J}{2C_r\omega},\; g_c = g_c^{\mathrm{static}} + \frac{\alpha C_r\omega}{2}.
\end{equation}
In general we will find that for a junction that works in the flux regime the term $\beta_J$ dominates all other contributions. But even without this assumption, it is true that while $g_c$ is fixed, the value of $g_i$ depends entirely on $\beta_J$ and can be changed in magnitude and sign, either enhancing the strength of the beam-splitter coupling~(\ref{eq:jc}), or switching it off entirely for a value of $\Phi_{\circlearrowright}\simeq \Phi_0/4$.

While the coupling strength grows with the loop size,  $\Delta{x}$, we can not make it arbitrarily large because then we are no longer allowed to neglect the additional flux $\phi_L$ caused by the circulating loop current due to the increasing value $L$ of  the self-inductance of the loop\cite{Orlando:1991a}. In this case the total flux threading the loop is given by the sum of the external flux $\Phi_{\rm ext}$ and the flux $\Phi_L$. However, as explained in Ref.~[\onlinecite{Orlando:1991a}] chapter 8.4, provided that
\begin{equation}
\beta_L = \frac{2\pi L I_c}{\Phi_0} < 1
\end{equation}
we can ensure that the $\Phi_\circlearrowright$ versus $\Phi_{\rm ext}$ dependence is single-valued allowing us full tunability of the coupling. This condition means that the maximum loop current $I_c$ cannot generate more than a single flux quantum. It restricts ourselves to loop sizes of around the $5\%$ of the resonator length. We will now study various methods to increase the coupling strength while preserving the condition above.

\subsection{Estimation of the coupling strength}
\label{subsec:strength}

\begin{figure}[t!]
  \centering
  \includegraphics[width=\linewidth]{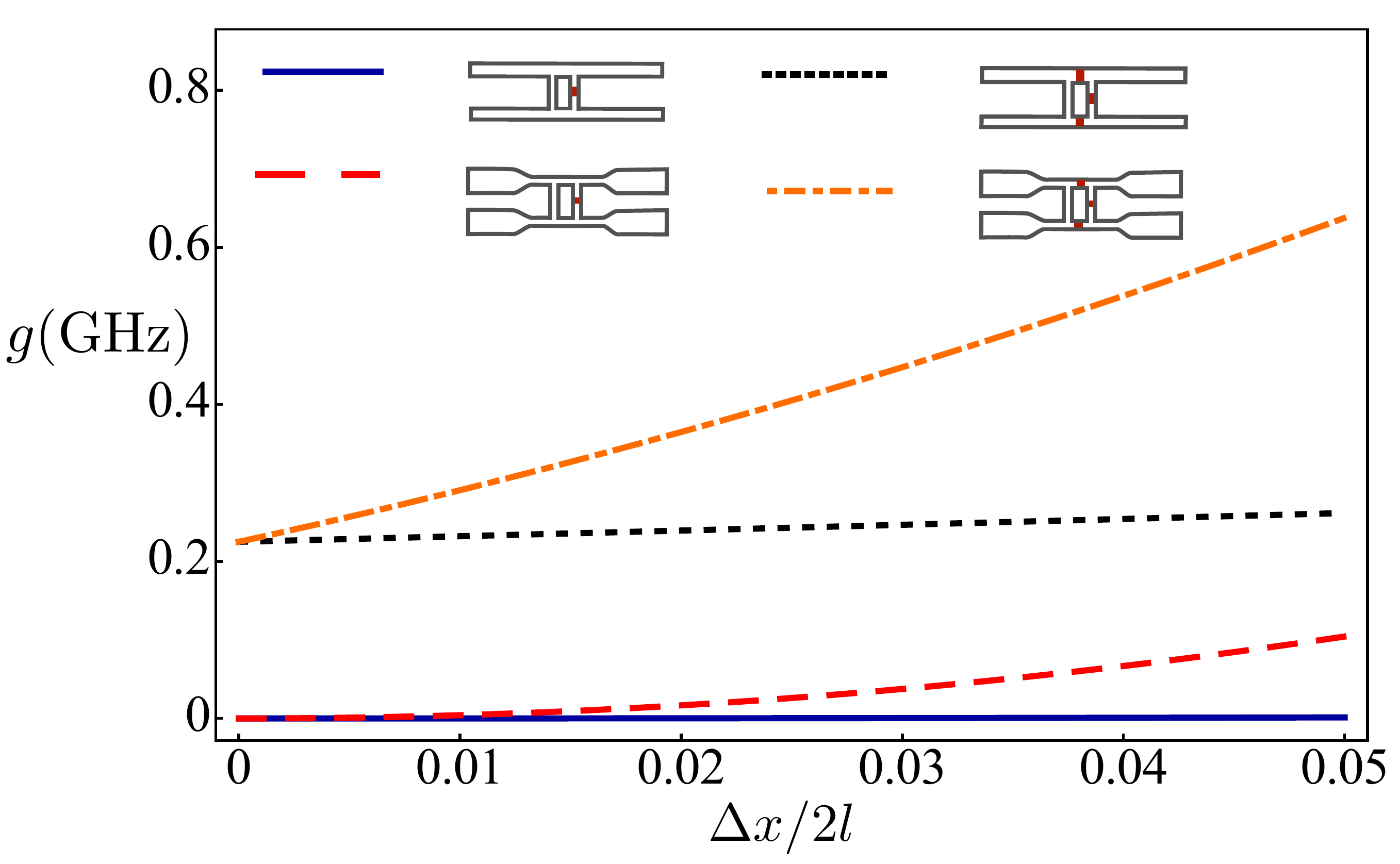}
  \caption{\small{Coupling strength for differerent Niobium
      transmission lines, as a function of the loop size. While homogeneous resonators (blue)  hardly reach the strong
      coupling regime, inhomogeneous ones (red-dashed) do. The coupling strength can be further increased with a Josephson junction interrupting the center conductor, as shown for the homogeneous case (dotted) and the inhomogeneous one (dot-dashed). We have considered for each resonator a frequency $\omega_0/2\pi=6.65\textrm{ GHz}$}.}
  \label{couplingrates}
\end{figure}

We are interested in an upper bound for the coupling strength
``$g$''. More precisely we would like to access both  the strong and
ultrastrong coupling regimes.
Strong coupling  means that it is possible to observe Rabi oscillations between both cavities because the coupling $g$ is  larger than the resonator decay rate, $\kappa$. 
On the other hand, ultrastrong coupling occurs when the RWA fails, which in this case implies that the number of photons in the ground state, that is proportional to $g/\omega$,  approaches one.

Looking at the first proposal [cf. Fig. \ref{setup_beam}a and Sect. \ref{sect:squid-coupler}]  we note that the maximum coupling is reached for an external flux $\Phi_{\circlearrowright}=n\Phi_0$ threading the SQUID loop, and thus yielding
\begin{equation}
\label{g_SQUID}
g_i\simeq\frac{4e^2}{2C_r}\frac{E_J}{\hbar^2 \omega}|u_1(x)|^2=\frac{I_c}{\Phi_0}Z|u_1(x)|^2,
\end{equation}
where $I_c$ is the critical current of the junction, $Z$ is the resonator impedance, and the eigenmode $u_0(x)$ satisfies $0<|u_1(x)|<\sqrt{2}$. To preserve the power field expansion, we suppose the SQUID to be built at a position such that $|u_1(x)|\leq 0.1$ [see Sect.~\ref{subsec:hopping}]. Under this condition, and using a critical current $I_c\simeq 5\times 10^{-6} \textrm{A}$, together with $Z=50~\Omega$, it would be possible to reach a coupling strength up to $g_{\textrm{i}}\simeq 1.2 \textrm{ GHz}$.

On the other hand,  for the second proposal [cf. Fig.~\ref{setup_beam}b and Sect. \ref{sect:ring} ] the coupling (\ref{coupling}) in the loop becomes
\begin{equation}
\label{g_slope}
g_{\textrm{i}}\simeq
  E_J\left (\frac{2 \pi}{\Phi_0}\frac{\partial\psi(x)}{\partial x}
  \Delta x\right )^2,
\end{equation}
with $\partial_x\psi=\partial_x u\sqrt{\frac{\hbar}{2C_r\omega}}$. For a
homogeneous resonator (see (\ref{ap:single})), we
can straightforwardly assess the slope of $u_1(x)$, finding
an exact expression for $g$:
\begin{equation}
\label{g_analitical}
g=\pi^2\frac{\omega_J\omega_c}{\omega_0}\left (\frac{\Delta x}{2l}\right )^2,
\end{equation}
where $\omega_J=E_J/\hbar$, $\omega_c=E_C/\hbar=(2e)^2/2\hbar C_r$ is the characteristic charging frequency of the resonator, and $\omega_0$ the first-mode frequency. Using available values for Nb striplines and junction parameters, we find that the homogeneous resonator remains in the weak coupling regime, as we envision before. For a loop size $\Delta x$ of $1\%$ of the resonator length, we obtain $g\simeq 2 \textrm{ MHz}$ which represents $0.03\%$ of the resonator frequency $\omega$ [see Fig.~\ref{couplingrates}]. 

Adding a constriction to the central part of the resonator increases the field slope and thus the coupling. To this end, for a suitable Nb inhomogeneous transmission line resonator~\cite{bourassa09},  this enhances the coupling up to $g\simeq 100 \textrm{ MHz}$, or equivalently to $1.8\%$ of the resonator frequency [Fig.~\ref{couplingrates} red-dashed].

Finally, the coupling can be further enhanced by interrupting the transmission line with a Josephson junction. Due to the presence of the junction, the flux eigenmode presents a constant phase slip $\Delta\phi_0$ at $x=0$, which depends on the Josepshon coupling energy of the junction\cite{bourassa09,bourassa12}. This additional phase jump enhances the coupling as follows
\begin{equation}
\label{g_junction}
g_{\textrm{i}}\simeq
  E_J\left (\frac{2 \pi}{\Phi_0}\frac{\partial\psi(x)}{\partial x}
  \Delta x+\Delta\phi_0\right )^2.
\end{equation}
Optimal parameters for the junction attached to the resonator ($E_{J\textrm{res}}\simeq 7E_{J\textrm{loop}}$) could lead to extremely large couplings of around $g\sim 600 \textrm{ MHz}$ ($9\%$ of $\omega$).

\subsection{Sidebands}
\label{sec:sidebands}

So far we have discussed the possibility of tuning the coupling strength between two resonators, constructing a classical switch that allows us to control the exchange of photons. In this section we discuss a second type of tunability, which consists of engineering an arbitrary linear coupling type between two resonators:
\begin{equation}
 H_{\mathrm{int}} = g_1 e^{i\phi_1} a^\dagger b + g_2 e^{i\phi_2} a b + \mathrm{H.c.},
\end{equation}
represented by the Fock operators $a$ and $b$. This would enlarge the applicability of our setup, extending it to the realization of almost any quadratic model with nearest neighbor interactions.

In order to demonstrate that this is possible we start our discussion by noting that both the dc-SQUID and the ring coupler provide us with a flux-dependent coupling
\begin{equation}
  H = \hbar\omega_a a^\dagger a+ \hbar\omega_b b^\dagger b +
  g(\Phi_{\circlearrowright})(a^\dagger + a )(b^\dagger + b).
\end{equation}
If we now engineer the two resonators to have very different frequencies, $\omega_a$ and $\omega_b$, a static coupling $|g|\ll \omega_{a,b}$ will be effectively suppressed, giving rise to a small dispersive term
\begin{equation}
  \label{H.free}
  H  \sim  \hbar\omega_a a^\dagger a+ \hbar\omega_b b^\dagger b +
  \frac{g^2}{|\omega_b-\omega_a|} a^\dagger a~b^\dagger b.
\end{equation}
However, if we allow for a two-tone driving of the coupling
\begin{eqnarray}
\label{g-t}
 g(t) = g[\Phi_{\circlearrowright}(t)] &=& g_1 \cos[(\omega_b-\omega_a)t+\phi_1] +\\
 &+& g_2 \cos[(\omega_a+\omega_b)t+\phi_2],\nonumber
\end{eqnarray}
then this driving effectively activates the rotating and counter-rotating terms, with the phases given above. To show this we switch to an interaction picture with respect to the two harmonic oscillators
\begin{equation}
  H_I = g(t) \left(a^\dagger b e^{i(\omega_a-\omega_b)t} + a b e^{-i(\omega_a+\omega_b)t} + \mathrm{H.c.}\right).
\end{equation}
The oscillating terms in Eq.~(\ref{g-t}) will precisely cancel the ones in the previous time-dependent Hamiltonian, leaving back some other non-resonant terms which only act in higher-order perturbation theory, $\mathcal{O}(g^2/\omega)$. The result should be the desired combination of sidebands
\begin{equation}
 H_{eff} = g_1 (a^\dagger b e^{i\phi_1} + a b e^{i\phi_2} + \mathrm{H.c.}) + \ldots
\end{equation}

It is worth mentioning that the previous sideband engineering is not perturbative: while we still need to impose that $|g_{1,2}|\ll |\omega_b-\omega_a|$, the resulting coupling is larger than the dispersive term. Another very important feature is that this method allows us to control the phase of the rotating and counterrotating terms, for this is related to the phase of the two-tone driving. As we discuss below, this is a very important property, as it allows us to implement effective gauge fields that control the hopping of photons between resonators. Moreover, we achieve this effect by a simple driving of a standard SQUID, without the need of time-reversal breaking circuits which might be very sensitive to other noise sources~\cite{nunnenkamp11}.

Finally, even though the realization of the time dependence~(\ref{g-t}) might seem very complicated, in practice we do not need to tune the flux in a very complicated manner. A simple driving of $\Phi_{\circlearrowright}(t) \simeq \Phi + \delta \Phi \cos(\omega t)$, when introduced in the sinusoidal coupling~(\ref{g-dependence}), $g \simeq \cos(2\pi\Phi_{\circlearrowright}/\Phi_0)$ produces, via the Jacobi-Anger expansion
\begin{equation}
  g(t) \simeq \cos(2\pi\Phi/\Phi_0) J_0(\delta \Phi) + \sin(2\pi\Phi/\Phi_0) J_1(\delta \Phi)\cos(\omega t) + \ldots
\end{equation}
in terms of the Bessel functions $J_0$ and $J_1$.
This series contains the basic driving plus higher harmonics which will be spectrally suppressed in the coupling term. Alternatively, a suitable dependence for $\Phi_{\circlearrowright}$ can be engineered with around 0.1 ns resolution using appropriate signal generators. Again, out of this signal only the resonant terms, with frequencies around $\omega_a\pm\omega_b$ will contribute to the coupling. Discretization errors in the signal, and higher harmonics, will be averaged out.

\subsection{Nonlinear photon hopping}
\label{subsec:hopping}
So far we have worked with the Josephson junctions in the linear regime, neglecting higher order terms, which are of the order $\frac{1}{24}E_J (2\pi\phi/\Phi_0)^4$. This approximation is valid only when the argument of the trigonometric functions, $2\pi/\phi/\Phi_0$, is small, a condition which can be recasted as a restriction on the number of photons that can populate the resonator. Roughly, for the SQUID we have the condition
\begin{equation}
 \phi \sim u(x) \sqrt{\frac{\hbar Z}{2} n} \ll \frac{\Phi_0}{2\pi} = \frac{\hbar}{2e},
\end{equation}
where $Z$ is the impedance of the resonators, $n$ is the average number of photons and $u(x)$ is the mode wavefunction at the coupling points. Using, in the same way as above, the value $u(x)=0.1$ restricts the number of photons to be $n < 1000$~[see App.~\ref{sec:linear}], which does not represent a restriction for the few photon applications that we envision.

The question now is what happens when we do not neglect the nonlinear terms. In this case we have the potential to introduce new interactions between resonators, which are now of higher order and include on-site nonlinearities, $n_i^2$, nearest-neighbor attractive or repulsive interactions, $n_in_j$, photon-pair hopping, $a^{\dagger 2}_ia^2_j$, etc [cf. App~\ref{sec:nonlinear}]. Out of these terms some are already strongly suppressed because of being off-resonant; this is the case for interactions with odd powers, such as $a^{\dagger 3}_ia_j$. The Kerr nonlinearities will always be present and give rise to extended Bose-Hubbard physics. Finally, the correlated hopping terms, $a^{\dagger 2}_ia^2_j$, can be resonantly enhanced using the same technique that we employ for the sidebands: introducing a frequency mismatch between neighboring cavities and driving with exactly the frequency which is needed to select this process, $2(\omega_i-\omega_j)$. With all these tools we envision the possibility of engineering very interesting models, such as a condensate of pairs of photons~\cite{eckholt08}, which are very hard to engineer in other systems.

\section{Applications and discussion}
\label{sec:conclusions}

\begin{figure}[t!]
  \centering
  \includegraphics[width=\linewidth]{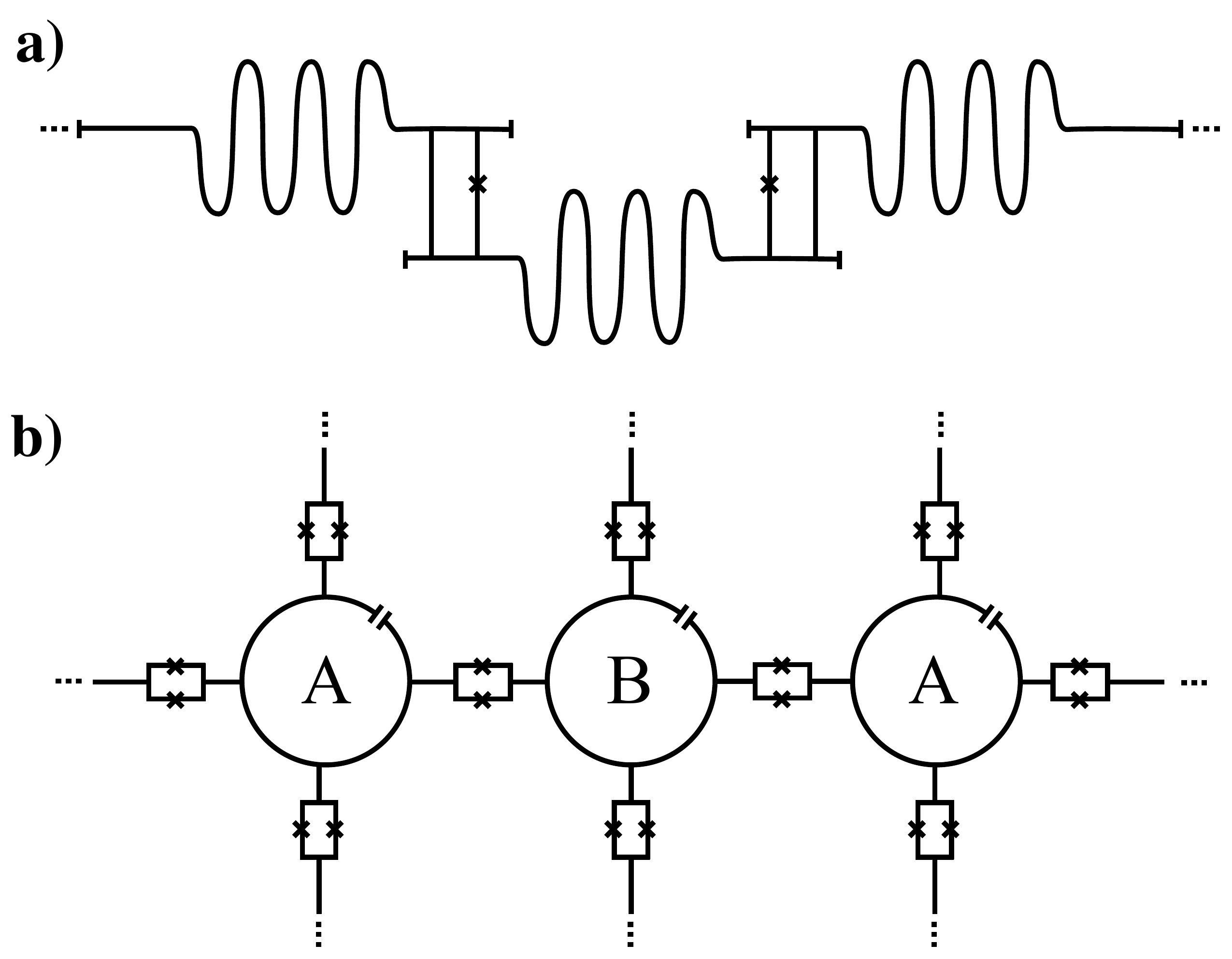}
  \caption{\small{a) One dimensional array of cavities coupled by means of a superconducting ring coupler. b) Two-dimensional lattice of circular resonators, coupled by dc-SQUIDs. Both lattices are bipartite. Using different resonator frequencies for each sublattice, $\omega_A \neq \omega_B$, we can use the techniques from Sect.~\ref{sec:sidebands} to engineer any sideband interaction between the arrays.}}
  \label{fig:array}
\end{figure}

Summing up, in this work we have studied two different ways to engineer the coupling between superconducting resonators: one is geometric and static in nature, while the other relies on nonlinear coupling circuits and can be easily tuned in and out of the strong coupling regime. Both elements together form a powerful toolbox for implementing almost arbitrary models consisting of a low-dimensional (from 1D to 2D) array of resonators with tunable nearest neighbor interactions, as in the model sketched in Fig.~\ref{fig:array}. Let us now discuss some of the potential applications of such circuits.

\paragraph{Traditional Quantum Optics.}

The implementation of tunable sidebands in coupled resonators opens the door to many well-known processes from quantum optics. Some of them are the squeezing of different modes via those sidebands, frequency conversion of photons as they are transferred between cavities, parametric generation of photons via $a^\dagger b^\dagger +ab$ terms, entanglement production at high temperatures \cite{Galve2010}, etc ... The beam-splitter Hamiltonian is also the cornerstone of all-optical quantum information processing, as suggested in Ref.~\onlinecite{chirolli10} for a different circuit-QED architecture. 

\paragraph{Harmonic models.}

The most immediate application of our design would be to implement arbitrary quasi-local and quadratic Hamiltonians, with the aims of studying the dynamical or statical properties of many-body Gaussian states. This includes a variety of studies, such as the static correlations in the model~\cite{Cramer06} and their relation to the underlying entanglement, dynamical quantum phase transitions from trivial to critical phases, the study of propagation of correlations in non-equilibrium models and their relation to Lieb-Robinson bounds~\cite{Eisert09,Cramer08,Plenio04}, etc. In this context, the tunability of the coupling plays two different roles. On the one hand it allows us to change the parameters of the Hamiltonian in an abrupt or smooth way, for instance to study a dynamical quantum phase transition or a quench. On the other hand and equally important,  by switching off all couplings we can freeze the quantum state of the oscillators, giving us time to measure the properties of the system, either with different measurement qubits or through a movable probe~\cite{shanks12}.

\paragraph{Anharmonicity \& thermalization.}

The interest of the harmonic problems lays in their simplicity and the possibility of obtaining analytical and numerical results for different geometries and sizes. However, as soon as we introduce a small nonlinearity in our system, we can say very little about their dynamical and static properties and many of the simulations which we mentioned in the previous paragraph become open problems. In particular, one very simple problem which deserves being studied is that of thermalization. The basic idea is to replace the linear resonators in Fig.~\ref{fig:array} with resonators that host a tunable and weak nonlinearity in the form of a SQUID (similar to Ref.~\onlinecite{Wilson11} but outside the linear regime). One would then prepare the ground state of the cavities with a value of the coupling, and then abruptly quench this coupling to a different (larger or smaller) value in which the prepared state is not a ground state. Throughout this process it will be possible to track the relaxation of the oscillator chain or lattice, studying how its behavior is modified by the presence of the nonlinearity.

\paragraph{Coupled cavities \& gauge fields}

Along the lines of anharmonic systems, another interesting problem is the study of coupled cavities or Jaynes-Cummings lattices~\cite{hartmann06,angelakis07}. The setup would be that of Fig.~\ref{fig:array}, but with one qubit attached to each resonator. The coupled qubit-resonator system behaves as a highly nonlinear element, implementing a quasiparticle known as ``polariton'', which may hop from resonator to resonator through our tunable coupling elements. This can be roughly formulated as a Bose-Hubbard Hamiltonian
\begin{equation}
  H = \sum_{ij} t_{ij} a_i^\dagger a_j + U(a^\dagger_ia_i),
\end{equation}
with a very nonlinear on-site interaction $U$ and a hopping $t_{ij}$ which, unlike previous proposals~\cite{leib10}, is now dynamically tunable. This allows us to explore the quantum phase transitions from weak interactions $U\ll |t|$ to hard-core particles $U\gg|t|$ simply by reducing the hopping instead of arbitrarily boosting the qubit-resonator interaction --- something which might be more challenging from the theoretical and experimental point of view.

In addition to the usual Mott-superfluid phase transition, we now have control over the phase of the hopping, $t_{ij} = |t| \exp{i\theta_{ij}}$. The procedure, as described in the previous section, consists of engineering two coupled sublattices [A and B in Fig.~\ref{fig:array}b] of resonators with different frequencies, $\omega_A \neq \omega_B$. Applying a multitone driving on the bonds that connect both sublattices, we can create an array of phases $\theta_{ij}$ which have a nontrivial flux around each plaquette. This will allow us to probe the integer quantum Hall physics with polaritons, without the use of circulators~\cite{nunnenkamp11}.

In summary, we have shown that in circuit-QED, tunable coupling between resonators can be implemented via simple Josephson circuits. We have developed this initial idea into a profound theoretical basis for exciting multi-resonator experiments ranging from arbitrary sideband interactions to setups scalable towards the many-body regime. On the theoretical side, our results lend themselves to be expanded to advanced scenarios, such as the relation between our circuit models and Josephson junction arrays, the influence of decoherence, or even for the design of models with tunably dissipative elements.

\acknowledgements

We thank Juan Jos\'e Mazo for useful discussions. This work was supported by EU projects PROMISCE and CCQED, Spanish MICINN Projects  FIS2009-10061, FIS2011-25167, and CAM research consortium QUITEMAD S2009-ESP-1594. B.P. acknowledges financial support from CSIC JAE-PREDOC2009 contract. F. W., F. D., A. M., and R. G. acknowlegde support by the German Research Foundation via SFB 631 and the German Excellence Initiative via the Nanosystems Initiative Munich (NIM). 
 
\appendix

\section{LUMPED ELEMENT MODEL OF TWO COUPLED RESONATORS}
\label{app:A}

Here we derive the density Lagrangian~(\ref{density_lag}) of Sect.~\ref{sec:inductive} from the quantum network theory perspective.

The appendix is divided into three parts: in the first one we review the quantization of a single microstrip resonator. In the second one, we consider the equivalent circuit of the coupled strip lines shown in Fig.~\ref{fig:lumped}a in its lumped element model [see Fig.~\ref{fig:lumped}b]. The Kirchhoff equations derived here will give rise to the Lagrangian~(\ref{density_lag}) in the continuum limit

\subsection{Single oscillator description}
\label{ap:single}
Here, we detail the description for the single resonator case.  The transmission line field equations are obtained from their lumped circuit equivalent.  Neglecting losses it can be described as a series of LC circuits\cite{Orlando:1991a}. In the continuum limit, the resulting field equations can be obtained from the Lagrangian:
\begin{equation}
{\mathcal L_0}=\int_{-l}^{l} d x \left[c_0 (x) \; \dot{\phi} (x,t)^2 \, - l_0(x)^{-1} \partial_x \phi(x,t)^2 \right],
\end{equation}
where $c_0(x)$ and $l_0(x)$ are the capacitance and inductance per unit of length, respectively; otherwise, $\phi(x)=(\Phi_0/2\pi) \varphi(x)$ is the magnetic flux variable with $\Phi_0 =h/2e$ the magnetic flux quantum and $\varphi(x)$ the phase of the macroscopic wavefunction describing the superconductor. The stationary modes are found by solving the eigenvalues and eigenvectors for the equation of motion (the Euler-Lagrange equations)
\begin{equation}
\label{lagrange}
\partial_x \big [
l_0(x)^{-1} \partial_x \phi (x, t)
\big ]
=
c_0(x) \partial_t ^2 \phi (x,t),
\end{equation}
which is nothing but the wave-equation in one dimension. The solution to this equation is expanded in normal modes and time dependent amplitudes,
\begin{equation}
\label{app:nmexp}
\phi(x,t) = \sum_n q_n(t) u_n(x),
\end{equation}
such that $\ddot q_n = -\omega_n q_n$, with $\omega_n$ the resonator
frequencies.  Therefore the {\it eigenstates} $u_n$ satisfy the
differential equation $\partial_x \big [ l_0(x)^{-1} \partial_x u_n
(x) \big ] = -\omega_n c_0(x) u_n(x) $. The $u_{ n}$ satisfy the
orthogonal relation:
\begin{equation}
\label{cr}
\int_{-l}^{l} c_{0,j} (x) u_{m,j}(x) u_{n,j}(x)  dx = C_r
\delta_{nm} ,
\end{equation}
with $C_r = \int_{-l}^{l} c_0(x) dx$ the total capacitance of
the resonator.

For homogeneous resonators, $l_0$ and $c_0$ are constant, and we obtain the well known case of equispaced eigenfrequencies $\nu_n = (2n -1)/2 l \sqrt{l_0 c_0}$ with $2l$ the length of the superconducting resonator and $u_n=\sqrt{2} \sin ((2n -1) \pi x /2l)$. 

\subsection{Two coupled oscillators}
\label{ap:coupled}
Consider the lumped element model depicted in
Figure~\ref{fig:lumped}b. The discrete modes of the electromagnetic field inside
the strips are described as arrays of $LC$ oscillators, together with
the mutual inductances and capacitances, representing the crosstalk. By applying the current conservation law at each node of the circuit,
we obtain a set of dynamical equations for the node fluxes $\phi_{j,n}$:
\begin{eqnarray}
&&\Delta x c_0 (n)\ddot \phi_{1, n} + \Delta x c_m (n) (\ddot \phi_{1, n} -  \ddot \phi_{2, n})=\\
&&\frac{l_0 (n)}{\Delta x}\frac{(\phi_{1, n-1} - \phi_{1, n})}{l_0 (n)^2 - l_m (n)^2} - \frac{l_0(n)}{\Delta x} \frac{(\phi_{1, n} - \phi_{1, n+1})}{l_0(n)^2 - l_m (n+1)^2} \nonumber\\
&&-\frac{l_m(n)}{\Delta x}\frac{  (\phi_{2, n-1} - \phi_{2, n})}{l_0 (n)^2 - l_m (n)^2}-\frac{l_m(n+1)}{\Delta x}\frac{(\phi_{2, n+1} - \phi_{2, n}) }{l_0(n)^2 - l_m (n+1)^2},\nonumber
\end{eqnarray}
and the equivalent equation for the second resonator. 
Above $c_0(n),l_0(n)$ are the capacitance and inductance per unit
length respectively, while $c_m(n),l_m(n)$ represent the mutual
capacitance and mutual inductance coefficients. Notice that, in
general, all these parameters are position dependent.

The former
equations of motion are nothing but the Euler-Lagrange equations
associated to the following Lagrangian
\begin{equation}
\label{general_lag}
L = T - V,
\end{equation}
with
\begin{align}
T &= \frac{\Delta x}{2} \sum_{n, j} c_0 \dot \phi_{j,n}^2 + c_m(n) ( \dot
\phi_{j,n} - \dot \phi_{j+1,n} )^2, 
\\ V &= \frac{1}{2\Delta x} \sum_{n, j}
\frac{l_0}{l_0^2 - l_m (n)^2} ( \phi_{j,n} - \phi_{j,n-1} )^2 
\\
&-
\frac{l_m(n)}{l_0^2 - l_m (n)^2} ( \phi_{j,n} - \phi_{j,n-1} )(\phi_{j+1,n} - \phi_{j+1,n-1} ). \nonumber
\end{align}
We can now take the continuum limit
$\Delta x \to 0$, which implies:
\begin{itemize}
\item $\displaystyle{\phi_{j,n}\to\phi_j (x)}$,
\item $\displaystyle{\frac{(\phi_{j,n} - \phi_{j,n-1})}{\Delta x}}\to\partial_x \phi_j(x)$,
\item $\Delta x\displaystyle{\sum_{n}}\to\int_{-l}^{l} \textrm{ d} x$.
\end{itemize} 
Hence, the Lagrangian~(\ref{general_lag})~ends up in Eq.~(\ref{density_lag}) in the main text that we rewrite here for completeness,
\begin{equation}
\label{eq:L}
{\mathcal L} = \frac{1}{2} \sum_{j=1,2} \int_{-l}^{l} {\rm d} x\left[{\hat c}_{ij} \dot{\phi}_i (x) \dot{\phi}_j (x) - \hat l ^{-1}_{ij} \partial_x{\phi_i(x)}\partial_x{\phi_j(x)}\right].
\end{equation}

The fluxes, $\phi_i(x),$ are thus coupled by the capacitance
$\hat{c}(x)$ and inductance $\hat{l}(x)$ matrices given in the main
text~(\ref{cl_matrices}). The diagonal terms of these matrices represent the single resonator
Lagrangian ${\mathcal L}_0$ derived in the previous section, that
depends on $l_0(x)$ and $c_0(x)$. On the other hand, the off-diagonal
contributions represent  the coupling Lagrangian ${\mathcal{L}_1}$,
described by the parameters $l_{jj} = l_0, l_{ij}= l_m, c_{jj}= c_0 + c_m$ and $c_{ij}= -c_m.$
\newline

\subsection{Generalization to more oscillators}
\label{ap:general}
We now show that the quantum description of two coupled resonators presented above can be generalized to the case of $N$ coupled resonators. We therefore extend the sum in~(\ref{eq:L}) to $N$
\begin{equation}
\label{eq:Lgen}
{\mathcal L} = \frac{1}{2} \sum_{i,j=1}^N \int_{-l}^{l} {\rm d} x  \left[{\hat c}_{ij} \dot{\phi}_i (x) \dot{\phi}_j (x) - \hat l ^{-1}_{ij} \partial_x{\phi_i(x)}\partial_x{\phi_j(x)}\right],
\end{equation}
where the  $\hat{c}(x)$ and $\hat{l}(x)$ are now given by $N\times N$
matrices with self capacities (self inductances) on the diagonal and
the mutual capacities (self inductances) on the off-diagonal.
Following the same procedure as above we restrict ourself to the fundamental modes, split off the single resonator Lagrangians and write the interaction part as
\begin{eqnarray}
L_1 &=&  \frac{1}{2} \sum_N \left( c_m  \Delta_1 \dot q_j^2 - \frac{l_m^2}{l_{\rm0} (l_{\rm
    0}^2 - l_m^2)}  \Delta_2 q_j^2 \right) \\
&+& \sum_{i=1}^{N-1} \left( -c_m \Delta_1 \dot q_i \dot
q_{i+1}  +  \frac{l_m}{l_0^2 - l_m^2} \Delta_2 q_i q_{i+1} \!\!\right),\nonumber
\end{eqnarray}
only taking into account nearest neighbor interaction. 
The Hamiltonian can finally be written as
\begin{eqnarray}
\label{eq:H_n}
H/\hbar&=&\sum_{j=1}^N
\omega {a}_j^\dag{a}_j^{}- \sum_{j=1}^{N-1} g_{c}
({a}_j^\dag-{a}_{j}^{}) ({a}_{j+1}^\dag-{a}_{j+1}^{})\nonumber\\
&-&\sum_{i=1}^{N-1}g_{i} ({a}_{j}^\dag+{a}_{j}^{})
({a}_{j+1}^\dag+{a}_{j+1}^{}), 
\end{eqnarray}
with $\omega$, $g_c$ and $g_i$ identical to the two resonator case.
In particular from the resulting total Hamiltonian $H$ the normal frequencies can be found, giving:
\begin{eqnarray}
\label{n-f}
\omega_- &=& \omega_0 \sqrt{\frac{1}{1 + 2 C}\left (1 + \frac{L}{\nu ( 1 - L)}\right )},\\
\omega_+ &=& \omega_0 \sqrt{1 - \frac{L}{\nu ( 1 + L)}}.\nonumber
\end{eqnarray}
We finally notice that by making $l_c \to l$ ( $\nu \to 1$), i.e. two straight parallel resonators, the formulas for the normal modes match the case of two coupled LC circuits.
\section{Linear and non-linear couplings}
\label{ap:B}
The motivation in this appendix is twofold. On the one hand 
we estimate the validity of the linear approximation; on the other hand, we explicitly compute the first non-linear corrections to the coupling.

We first expand the cosine in ${\mathcal L}_{\mathrm{SQUID}}$ (\ref{LSQUID}),
\begin{equation}
\label{exp-1}
- \cos \left (
\frac{2 \pi} {\Phi_0} \phi_-
\right )
=
-1 +\frac{1}{2} \left (
\frac{2 \pi} {\Phi_0} \right )^2 \phi_-^2
-
\frac{1}{24} \left (
\frac{2 \pi} {\Phi_0} \right )^4 \phi_-^4
+ ...
\end{equation}
We recall that:
\begin{equation}
\label{phi-}
\phi_- =  u_0 \sqrt{ \hbar Z}  (a_-^\dagger + a_-)
\; ,\mbox{ and }
a_- = \frac{1}{\sqrt{2}} (a_1 - a_2)
\end{equation}
Assuming $Z=50\,\Omega$ and defining 
\begin{equation}
\label{xi}
\xi \equiv
u_0^2 \left (
\frac{2 \pi} {\Phi_0} \right )^2  2 \hbar Z
= u_0^2 \frac{\pi^2}{10^2} \cong 10^{-1} u_0^2
\end{equation}
we can write for the expansion:
\begin{equation}
\label{exp-2}
- \cos \left (
\frac{2 \pi} {\Phi_0} \phi_-
\right )
=
-1 +\frac{\xi}{4} (a_-^\dagger + a_-)^2
-
\frac{\xi^2}{96} (a_-^\dagger + a_-)^4
\end{equation}

\subsection{ Linear regime}
\label{sec:linear}

The linear approximation is justified when the second order terms in Eq.~(\ref{exp-1}), or equivalently the average value and fluctuations of the flux in Eq.~(\ref{exp-2}), are small. How does this relate to actual experiments? In order to determine a condition based on the number of photons we study the fluctuations $\langle \phi_-^2\rangle$, which are related to the expectation value
\begin{equation}
\langle  (a_-^\dagger + a_-)^2 \rangle 
\cong 2 ( \langle a_1^\dagger a_1 \rangle + \langle a_2^\dagger a_2 \rangle)
\equiv 4 n
\end{equation}
with $n$ the number of photons. Using the previous series we conclude that linearization is strictly justified whenever $n \ll 10 / u_0^2$, where $u_0$ is the value of the mode wavefunction.
For the ring coupler  layout, the same reasoning follows by replacing $u_0 \to \partial_x u_0 \Delta x$ in~(\ref{phi-}) and~(\ref{xi}).
\subsection{Non-linear hopping terms}
\label{sec:nonlinear}
 
With the help of Pathak's results \cite{Pathak2000}, we compute,
\begin{equation}
(a_- + a_-^\dagger)^4 = a_- ^4
+
4 (a_- ^\dagger)^3 a_- 
+
6  (a_- ^\dagger)^2 a_- ^2
+
6 a^\dagger a
+ 3 + \mathrm{H.c.}
\end{equation}
Notice that $a_- = \frac{1}{\sqrt{2}} (a_1 - a_2)$ [Eq. (\ref{phi-})], so the above ends up in a long expression hard to deal with. To make it simpler, we assume that sidebands will select $a_-^2$ or $a^\dagger a$ and we resort to a RWA argument to write:
\begin{equation}
(a_- + a_-^\dagger)^4
\cong
6  (a_- ^\dagger)^2 a_- ^2+12 a^\dagger a+6 (a_- ^\dagger)^2 +  6 a_- ^2+ 3.
\end{equation}
Looking at each term we rewrite in terms of the {\it local} bosonic operators, $a_1$ and $a_2$:
\begin{eqnarray}
4(a_- ^\dagger)^2 a_- ^2
&=& (a_1^\dagger)^2 a_1 ^2+(a_2^\dagger)^2 a_2 ^2+ 2 a_1 ^\dagger a_2^\dagger a_1 a_2 \\
{}&& - 2 a_1^\dagger a_2^\dagger \left(a_1^2 + a_2^2 \right) + \mathrm{H.c.}\nonumber
\\
2(a_- ^\dagger)^2 & =& (a_1^\dagger)^2 + (a_2^\dagger)^2 - 2 a_1^\dagger a_2^\dagger
\\
2a_-^\dagger a_-  & = &a_1^\dagger a_1+a_2^\dagger a_2-a_1^\dagger a_2-a_2^\dagger a_1.
\end{eqnarray}


\bibliographystyle{apsrev}

 \end{document}